\documentstyle[aps,graphicx,twocolumn]{revtex}

\newcommand{\half}{\mbox{\small\(\frac{1}{2}\)} }
\renewcommand{\d}{d} 
\newcommand{\ij}{i}  
\newcommand{\calE}{{\mathcal{E}}}
\newcommand{\calN}{{\mathcal{N}}}

\newcommand{\re}{{\mathrm{Re}}\,}
\newcommand{\im}{{\mathrm{Im}}\,}
\newcommand{\eq}[1]{Eq.~(\ref{#1})}
\newcommand{\fig}[1]{Fig.~\ref{#1}}
\newcommand{\eqs}{Eqs.}

\newcommand{\const}{{\mathrm{const}}}

\newcommand{\sech}{{\mathrm{sech}}\,}

\newcommand{\Eosc}{\widetilde{{\mathbf{E}}}}

\newcommand{\Av}{{\mathbf{A}}}
\newcommand{\Bv}{{\mathbf{B}}}
\newcommand{\ev}{{\mathbf{e}}}
\newcommand{\Ev}{{\mathbf{E}}}
\newcommand{\pv}{{\mathbf{p}}}
\newcommand{\vv}{{\mathbf{v}}}
\newcommand{\xv}{{\mathbf{x}}}

\begin{document}
\draft
\author{R. L. Dewar\thanks{Electronic address: robert.dewar@anu.edu.au}}
\address{
  Department of Theoretical Physics \& Plasma Research Laboratory,
  Research School of Physical Sciences \& Engineering,
  The Australian National University,
  Canberra, A.C.T. 0200, Australia}
\author{C. I. Ciubotariu}
\address{Physics Department, 
  University of Lethbridge, Lethbridge T1K 3M4, Alberta, Canada}
\date{\today}
\title{Quasilinear theory of collisionless Fermi acceleration in a multicusp
magnetic confinement geometry}
\maketitle

\begin{abstract}
Particle motion in a cylindrical multiple-cusp magnetic field 
configuration is shown to be highly (though not completely) chaotic, 
as expected by analogy with the Sinai billiard.  This provides a 
collisionless, linear mechanism for phase randomization during 
monochromatic wave heating.  A general quasilinear theory of 
collisionless energy diffusion is developed for particles with a 
Hamiltonian of the form $H_0+H_1$, motion in the \emph{unperturbed} 
Hamiltonian $H_0$ being assumed chaotic, while the perturbation $H_1$ 
can be coherent (i.e. not stochastic).  For the multicusp geometry, 
two heating mechanisms are identified --- cyclotron resonance heating 
of particles temporarily mirror-trapped in the cusps, and nonresonant 
heating of nonadiabatically reflected particles (the majority).  An 
analytically solvable model leads to an expression for a transit-time 
correction factor, exponentially decreasing with increasing frequency.  
The theory is illustrated using the geometry of a typical laboratory 
experiment.
\end{abstract}


\section{Introduction}

The quasilinear diffusion equation in its original form (QLT1) was a 
Fokker-Planck equation describing the velocity-space diffusion of 
particles due to random scattering by waves.  In the absence of the 
waves the plasma was assumed to be infinite and homogeneous so that 
the unperturbed motion was rectilinear.  The diffusion equation was 
derived \cite{vedenov-velikhov-sagdeev62,vedenov63,drummond-pines62} 
from the Vlasov equation by solving for the perturbed part of the 
distribution function, in the linear approximation and assuming the 
unperturbed distribution function to be essentially constant over the 
timescale of wave-particle interaction, then substituting back into 
the full Vlasov equation and averaging over position (or, 
equivalently, over the random phases of the waves).

To put this formalism in a more general perspective, it is 
advantageous to cast it in Hamiltonian form, with the particle motion 
in the absence of waves being described by an unperturbed Hamiltonian 
$H_0$.  The total single-particle Hamiltonian is then $H_0+\epsilon 
H_1$, with the waves being incorporated in the perturbation 
Hamiltonian $H_1$.  In QLT1, all stochasticity comes from the 
assumption of random phases in the assumed broad spectrum of waves in 
$H_1$.  The smallness parameter $\epsilon$ expresses the assumption 
that the amplitudes of individual waves are small, so that the 
short-time-scale perturbations to the distribution function can be 
derived using linear, $O(\epsilon),$ theory (hence the terminology 
``quasilinear'').  The only nonlinear effect is the long-time 
diffusion described by the diffusion coefficient, which is 
$O(\epsilon^2)$ .

It is not really necessary to assume an infinite, homogeneous plasma.  
The essence of QLT1 is the assumption that $H_0$ is \emph{integrable}, 
so that a canonical transformation to action-angle coordinates exists.  
Then the unperturbed motion in angle space is rectilinear, just as that 
in coordinate space in the case of a homogeneous medium.  This 
action-angle generalization was carried out by Kaufman 
\cite{kaufman72} and by Hazeltine \emph{et al.} 
\cite{hazeltine-mahajan-hitchcock81} to derive a quasilinear diffusion 
equation (in action space) for wave-particle scattering in 
axisymmetric toroidal magnetic confinement geometries (e.g. tokamaks).  
The formalism has been used, for example \cite{shurygin-dewar95}, to 
investigate the effect of a sheared radial electric field on anomalous 
transport in a tokamak.

With the development of the theory of Hamiltonian chaos it has come to 
be realized that a quasilinear diffusion equation can also be derived 
in cases where $H_1$ represents the effect of a \emph{coherent} wave, 
provided the interaction of $H_0$ and $H_1$ produces a chaotic motion.  
We call this form of quasilinear theory QLT2, and it is very useful in 
the theory of radio-frequency (rf) and microwave heating of plasmas 
\cite{stix92} because this is typically done with coherent waves.

Again, $H_0$ is assumed integrable so that an action-angle 
transformation exists.  In these theories the \emph{perturbation} is 
still the source of chaos (``intrinsic stochasticity''), which causes 
the action variables, constructed in the integration of $H_0$, to 
perform the random walk described by the quasilinear diffusion 
equation.  Assuming the Hessian matrix $\partial^2 H_0/\partial 
p_i\partial p_j$ to be nonsingular, the coherent perturbation must 
exceed a certain amplitude for global resonance overlap, and hence 
chaos, to occur 
\cite{chirikov79,greene79,rechester-rosenbluth-white79}.  Thus, 
paradoxically, a criterion for QLT2 to apply is that the system be 
sufficiently nonlinear!

In the present paper we examine a third form of quasilinear theory, 
which we call QLT3.  This case is the complete obverse of the original 
quasilinear diffusion problem: we now assume the unperturbed 
Hamiltonian $H_0$ to be completely nonintegrable, giving rise to 
strongly chaotic motion in the \emph{unperturbed} system.  The chaotic 
dynamics of the unperturbed Hamiltonian system then provides 
randomization and allows the application of a quasilinear formalism to 
derive the diffusion equation for the distribution function.

Of course, since $H_0$ is not integrable, an action-angle 
representation does not exist.  In fact we assume $H_0$ to be so 
strongly chaotic that the unperturbed motion covers essentially the 
entire energy surface $H_0 = \calE$ ergodically, except as restricted 
by integrals of the motion associated with any continuous symmetries 
of $H_0$.  The goal of this paper is to determine the diffusion in 
$\calE$ caused by the time-dependent perturbation $H_1$.

Since the unperturbed motion provides the source of stochasticity 
(with no threshold), we can, as in QLT2, assume the perturbation to be 
coherent.  Thus the theory is applicable to wave heating of plasmas 
in strongly nonintegrable magnetic confinement geometries.

The assumption of uncorrelated gyrophase in successive passes through 
the resonance region is often made in the derivation of a quasilinear 
diffusion equation to describe cyclotron resonance heating of 
magnetically confined plasmas.  However, in simple confinement 
geometries, such as magnetic mirrors, $H_0$ is essentially integrable 
owing to the existence of the adiabatic invariant $\mu$ and another 
integral due to symmetry, or a second adiabatic invariant.  
Lichtenberg and Lieberman \cite{lichtenberg-lieberman92} have analyzed 
collisionless heating in such systems using area-preserving maps, and 
find the random phase assumption to be valid only well beyond the 
nonlinear threshold where the last invariant circle is destroyed and 
chaotic motion becomes global.  In our nomenclature, quasilinear 
diffusion in these systems is an example of QLT2, not QLT3.

On the other hand, in systems with a null in the magnetic field $\mu$ 
is not globally conserved and the situation is rather different from 
that in the much-studied mirror systems.  For instance, Yoshida 
\emph{et al.} \cite{yoshida-etal98} have recently studied rf heating 
in a two-dimensional slab model with a neutral line, where $\mu$ 
conservation is broken.  They find heating due to the onset of chaos 
at finite amplitude of the perturbing field.  However, their 
unperturbed system has two symmetry directions, and thus their $H_0$ 
is integrable, despite the breaking of the adiabatic invariant.  Thus 
their model must be classified as a QLT2 case also.

In systems with a null in the magnetic field and only \emph{one} 
symmetry direction, however, $H_0$ is not in general integrable.  An 
important class of such systems are the magnetic cusp confinement 
geometries, which are much used in low-temperature plasma physics due 
to the ease with which they can be created with arrays of permanent 
magnets \cite{burke-pelletier92}\cite[pp.  
146--150]{lieberman-lichtenberg94}.  In this paper we give evidence 
that they fulfill the criterion for systems of type QLT3 in that their 
unperturbed dynamics is almost completely chaotic.

We know from the work of Sinai \cite{sinai70,ott93} that particle 
motion on a billiard table with a convex boundary is a strongly 
chaotic system due to the defocussing effect of each collision with 
this boundary.  In fact Sinai showed the motion to be strongly mixing, 
so that ergodic theory could be applied.  This suggests that cusp 
confinement systems, where magnetic fields lie on surfaces that are 
convex toward the plasma, are strongly chaotic.  Indeed the four-cusp 
Hamiltonian $H = \half(p_x^2 + p_y^2 + x^2 y^2)$ was at one time 
conjectured to be completely chaotic like a Sinai billiard.  Although 
Dahlqvist and Russberg \cite{dahlqvist-russberg90} disproved this 
conjecture by finding a stable periodic orbit, they found that the 
island of stability surrounding this orbit was extremely small, so for 
practical purposes one can assume that the energetically accessible 
phase space is covered ergodically in this system.

In this paper we study a magnetic field configuration consisting of a 
``picket fence'' of infinite linear magnetic dipoles, producing 
multiple line cusps.  This can be regarded as a simplified model for a 
low-temperature plasma source, created using arrays of permanent 
magnets \cite{ciubotariu-golovanivsky-bacal96,ciubotariu97}.  It is 
also a rather simple model, in which the unperturbed dynamics can be 
simplified analytically to a considerable extent by the use of complex 
variable theory.

In the limit of a large number of dipoles the interior of the plasma 
is essentially free of magnetic field and the unperturbed motion is 
rectilinear in the interior region, while near the edge of the 
confinement region, a particle can be reflected over a range of 
angles.  Thus we might expect the configuration to approximate a 
chaotic billiard problem.  (In the original Sinai problem the convex 
boundary was in the interior of the domain, whereas the billiard 
analog of the present example has an outer boundary that is convex 
except for cusps, like the ``bow-tie'' billiard shown in Fig.~7.24(e) 
of Ref.~\cite{ott93}.)

The problem is also related to a model originally proposed by Fermi 
\cite{fermi49,jarzynski94,lieberman-godyak98} for explaining the 
acceleration of cosmic rays to the extraordinarily high energies 
observed.  In the Fermi model the cosmic ray particles move 
rectilinearly except during occasional collisions with moving 
magnetized clouds of gas, which cause diffusion in energy space.  The 
present model includes both the possibility of cyclotron-resonance 
heating in the mirror-like cusp regions and nonresonant heating of 
particles reflected without penetrating deeply into the cusps.  The 
latter case is much closer to the original Fermi problem and is the 
main focus of the paper.

In Sec.~\ref{sec:2D} we introduce the confinement geometry and 
unperturbed Hamiltonian in detail, and in Sec.~\ref{sec:H0dynamics} we 
analyze the dynamics of this system and show it is indeed strongly 
chaotic for the class of particles (``free particles'') that traverse 
the central low-field region.  However, in Sec.~\ref{sec:Perorb} we 
produce a counter example to any conjecture that the motion is totally 
chaotic by finding a stable periodic orbit.

In Sec.~\ref{sec:Wave} we introduce the wave-particle interaction.  
The general quasilinear diffusion equation is derived in 
Sec.~\ref{sec:Transport}.  An analytically solvable one-dimensional 
model of the magnetic field is used in Sec.~\ref{sec:Simple} to 
estimate heating of nonadiabatically reflected particles.  The result 
is of the form expected from simple Fermi acceleration theory 
multiplied by a transit-time reduction factor that becomes 
exponentially small when the transit time is much longer than the 
period of the applied field.  The theory is discussed using typical 
parameters for permanent-magnet confinement experiments in 
Sec.~\ref{sec:Discussion}.

\section{Unperturbed Hamiltonian}
\label{sec:2D}

\subsection{Two-dimensional magnetic Hamiltonian}
\label{sec:2DHam}

The behavior of a sufficiently dilute plasma can be analyzed on the 
basis of single-particle motion in magnetic and electric fields made 
up of an externally imposed component and an internally generated 
component produced by the collective currents and charges from the 
combined effect of many otherwise noninteracting particles.  In this 
paper we suppose that the self-consistent component is negligible and 
analyze single-particle motion in an imposed magnetic field.

Consider the motion of a particle of charge $q$ and mass $m$ in a 
straight, infinitely long magnetic confinement system with vector 
potential $\Av = \psi(x,y)\ev_{z}$, where $x$, $y$ and $z$ are 
Cartesian coordinates and $\ev_{z}$ is the unit vector in the 
$z$-direction.  The magnetic field is $B_{x} = \partial \psi/\partial 
y$, $B_{y} = -\partial \psi/\partial x$, $B_{z} = 0$, so contours of 
the flux function $\psi(x,y)$ in any plane $z = \const$ define 
magnetic field lines.

The Hamiltonian is (see e.g. \cite{schmidt66})
\begin{equation}
  H_0 =  \frac{p_{x}^{2}}{2m} + \frac{p_{y}^{2}}{2m}
   	+ \frac{[p_{z}-q\psi(x,y)]^{2}}{2m} \;,
    \label{eq:H0}
\end{equation}
where $p_i$ ($i \in \{x,y,z\}$) are the canonical momenta, Hamilton's
equations of motion being $\dot{x}_i = \partial H_0/\partial p_i$, 
$\dot{p}_i = -\partial H_0/\partial x_i$.

\subsection{Multicusp flux function}
\label{sec:Multiflux}

Assuming no currents present in the plasma, $\psi$ obeys Laplace's 
equation.  It is a standard result that two-dimensional solutions of 
Laplace's equation can be constructed by taking the real or imaginary 
part of any analytic function of $\zeta \equiv x + \ij y$ 
\cite{morse-feshbach53}.  Thus we write
\begin{equation}
		\psi(x,y) = \re\Psi(\zeta) \;,
		\label{eq:PsiDef}
\end{equation}
where $\Psi(x,y)$ is the \emph{complex flux function}.

For instance, a line current (line magnetic monopole) at $\zeta_{0}$ 
is represented by the real (imaginary) part of $\ln 
(\zeta-\zeta_{0})$.

Although a magnetic monopole cannot be realized in nature, a magnet 
can be modeled as a superposition of magnetic dipoles.  We consider a 
magnet that is long in the $z$-direction, thin in the $x$- and 
$y$-directions, and magnetized in the $x$-direction so that it can be 
modeled by a line magnetic dipole.  Such a linear dipole can be 
produced by differentiation of $\im\ln(\zeta-\zeta_{0})$ with respect 
to $x_{0}$.

Superimposing the flux functions for $2n$ linear magnets of strength 
alternately $+K$ and $-K$ lying in a circular cylinder of radius $a$ 
about the $z$-axis we find $\psi$ for a circular multidipole magnetic 
confinement system,
  \begin{eqnarray}
    \psi  & = & K\,\im\sum_{n' = 0}^{n-1}
    \left[
      \frac{u_{n}^{4n'+1}}
        {\zeta - au_{n}^{4n'+1}}
      - \frac{u_{n}^{4n'-1}}
          {\zeta - au_{n}^{4n'-1}}
    \right] \;, \nonumber \\
    & = & \frac{2nK}{a}\re\,
    \left[
      \left(\frac{\zeta}{a}\right)^{n} + \left(\frac{a}{\zeta}\right)^{n}
    \right]^{-1} \;.
    \label{eq:PF}
  \end{eqnarray}
Here $u_{n} \equiv \exp (\pi\ij/2n)$ is the $4n$th root of 
unity.  The equivalence of the first and second forms can be verified 
by showing that they have the same poles and residues and the 
same value at $\zeta = 0$.

\begin{figure}[tbp]
		\centering
		\includegraphics[scale=0.5]{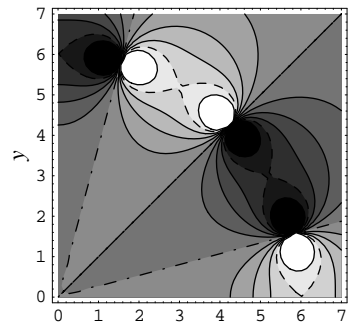}
		\caption{Contours of the flux function for a typical 
		cylindrical multicusp geometry produced by 12 line dipoles of 
		alternating polarity.}
		\label{fig:Contours}
\end{figure}

Thus, comparing with \eq{eq:PsiDef}, we see that the complex stream 
function for a circular dipole picket fence is given by
  \begin{eqnarray}
    \Psi & = & \frac{2nK}{a}
    \left[
      \left(\frac{\zeta}{a}\right)^{n} + \left(\frac{a}{\zeta}\right)^{n}
    \right]^{-1} \nonumber \\
		& \equiv & 
		\frac{nK}{a}\,\sech\!
			\left[
				n\ln \left(\frac{\zeta}{a}\right)
			\right]
		\;.
    \label{eq:PFcmplx}
  \end{eqnarray}
It is clear from the first form that $\Psi$ has poles at the $2n$th 
roots of $-a^{2n}$.  In terms of polar coordinates $r$ and $\theta$ 
such that $\zeta = r\exp\ij\theta$, the poles are at $r = a$, $\theta 
= \pi/2n + 2l\pi$, $l = 0,\,1,\ldots,2n-1$.  Contours of $\psi$ (lines 
of force) are shown in \fig{fig:Contours} for the case $n=6$, with 
distances measured in units of $a/n$.

\subsection{Near field and nondimensionalization}
\label{sec:Nearfield}

We now consider the behavior of $\psi$ in the region between two 
magnets, which for definiteness we take to be those immediately above 
and below the positive real $x$-axis.  We expand about the point 
$\zeta = a$, on the circle on which the magnets are located, by 
setting $\zeta = a+\xi$, where $\xi \equiv x-a + \ij y$.  Assuming 
$|\xi| \ll a$, we see from \eq{eq:PFcmplx} that
  \begin{eqnarray}
    \Psi \approx \frac{nK}{a}\,\sech\!
    \left(
      \frac{n\xi}{a}
    \right) \;.
    \label{eq:PFapprox}
  \end{eqnarray}
Thus, on the $x$-axis $\psi$ peaks at $x = a$ and it decays rapidly on 
either side.  Furthermore, we see that the scale length of the $x$ and 
$y$ variation is not $a$ but $a/n$.  In Sec.~\ref{sec:Scattering} we 
shall encounter \eq{eq:PFapprox} again in the context of the 
asymptotic limit $n \rightarrow \infty$.

Expanding $\Psi$ to second order in $n\xi/a$ we find $\psi \equiv \re 
\Psi \approx \psi_{\mathrm{X}}\{1 - n^{2}[(x-a)^{2} - y^{2}]/2a^{2}\}$ 
in the neighborhood of $(a,0)$, where
\begin{equation}
		\psi_{\mathrm{X}} \equiv nK/a
		\label{eq:psiXdef}
\end{equation}
is the value of $\psi$ at this saddle point, the location of a 
magnetic-field null. It is also useful to define a typical magnetic 
field $B_0$ in the strong-field region via the relation $B_0 \equiv 
n\psi_{\mathrm{X}}/a$.

We term the energy of a particle with momentum $\pv=0$ located at 
this saddle point the \emph{escape energy}
\begin{equation}
		\calE_{\mathrm{esc}} \equiv 
		\frac{q^2\psi_{\mathrm{X}}^{2}}{2m} \equiv 
		\frac{1}{2m}\left(\frac{qnK}{a}\right)^2 \;.
		\label{eq:Eescdef}
\end{equation}

In all numerical work and figures in this paper we nondimensionalize 
by measuring distance in units of $a/n$, mass in units of $m$, time in 
units of the typical inverse angular cyclotron frequency
\begin{equation}
		\tau_{\mathrm{X}}
		\equiv \frac{m}{|q|B_0}
		\equiv \frac{a^2m}{n^2Kq}\;,
		\label{eq:tUnit}
\end{equation}
and charge in units of $|q|$.  In these units $a=n$, 
$m = |q| = \tau_{\mathrm{X}} = \psi_{\mathrm{X}} = B_0 = 1$, and
$\calE_{\mathrm{esc}} = \half$.

\section{Unperturbed particle dynamics}
\label{sec:H0dynamics}

\subsection{Dynamics in complex notation}
\label{sec:Complexdyn}

For a given $p_z$, the dynamics is a two-degree-of-freedom 
autonomous Hamiltonian system which can be compactly written using 
complex notation as
\begin{eqnarray}
	\dot{\zeta} & = & \frac{1}{m}\, p_{\zeta} \;,\\ \nonumber
	\dot{p}_{\zeta} 
	& = & \frac{q}{m}
	\left[p_z - q\re\Psi(\zeta)\right][\Psi'(\zeta)]^{*}
	\label{eq:complexHeqns} \;,
\end{eqnarray}
where the prime on $\Psi$ means the derivative with respect to its 
argument, $*$ means complex conjugate, and $p_{\zeta} \equiv p_{x}+\ij 
p_{y}$.

\subsection{Integrability and effective potential}
\label{sec:Effpot}

Because $H_0$ is independent of $z$, $p_z$ is a constant of the 
motion.  Also, $H_0$ itself is a constant of the motion, $H_0 = 
\calE$, where the constant $\calE$ is the total energy.  However, the 
absence of a third integral of the motion means the system is not 
integrable.  Thus we must resort to numerical integration to 
investigate the nature of the unperturbed orbits.

Before proceeding to a discussion of the numerical results, however, 
we observe that some qualitative understanding of the motion can be 
found by determining the bounds of the motion implied by the constancy 
of $H_0$.  Inspecting \eq{eq:H0} we see that the term 
$V_{\mathrm{eff}} \equiv (p_z - q\psi)^2/2m$ in $H_0$ acts as an 
\emph{effective potential} in which the particles move.  The motion in 
the $(x,y)$-plane is thus bounded by the curves $V_{\mathrm{eff}}(x,y) 
= \calE$.  Note that $V_{\mathrm{eff}} \geq 0$, with equality 
occurring on the contours $\psi= p_z/q$.  Also, since $\psi$ 
vanishes at the origin, $V_{\mathrm{eff}} = p_z^2/2m$ there.

In the case $p_z = 0$, the curves $V_{\mathrm{eff}}(x,y) = \const$ are 
just level curves of $|\psi|$.  There are thus two cases:
\begin{enumerate}
	\item{Perfect confinement}\newline
		$\calE < \calE_{\mathrm{esc}}$, the curve $|\psi| = 
		(2m\calE)^{1/2}/|q|$ encloses the origin (though it has 
		cusps at the magnets) and there is thus no leakage of particles 
		with $p_{z} = 0$ through the dipole picket fence.
	\item{Partial confinement}\newline
		For $\calE > \calE_{\mathrm{esc}}$, the curves 
		$|\psi| = (2m\calE)^{1/2}/|q|$ are disjoint and thus 
		particles can escape through the ``mountain passes'' (see 
		\fig{fig:Contours}) between the magnets.
\end{enumerate}
In either case the particles are free to traverse a large region 
including the origin, like particles rolling on a billiard table [in 
case (ii) it is a billiard table with pockets].  We refer to particles 
on such orbits as \emph{free particles}, to be contrasted with the 
trapped particles discussed in the next section.

\begin{figure}[btp]
		\centering
		\includegraphics[scale=0.5]{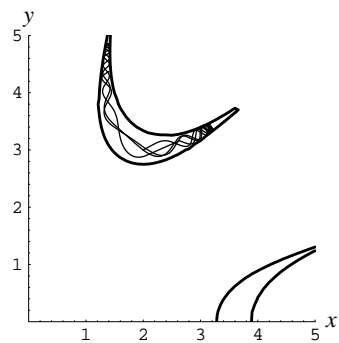}
		\caption{A typical trapped-particle orbit in the magnetic field 
		shown in \protect\fig{fig:Contours}.  The thick lines show the 
		boundaries of the energetically accessible regions.}
		\label{fig:MirrorOrbit}
\end{figure}

\subsection{Trapped-particles}
\label{sec:Traporb}

For $p_z \neq 0$ a new class of orbit arises, the \emph{trapped 
particles}.  This occurs when the energy is less than the value of 
$V_{\mathrm{eff}}$ at the origin, i.e. when $\calE < p_z^2/2m$, for 
then the low-magnetic-field central region is energetically 
inaccessible (see Sec.~\ref{sec:Freeorb}) and the particle must be 
confined in the edge region near the magnets, as illustrated in 
\fig{fig:MirrorOrbit}.

Because deeply trapped particles are always in a region of strong 
magnetic field, their dynamics can be analyzed (see, e.g., 
\cite[pp.  21--22]{schmidt66}) by decomposing the motion into that of 
the guiding center, with velocity $\vv = v_{\parallel}\Bv/B + 
\mathrm{drifts}$, and a gyro motion with velocity $\vv_{\perp}$ in 
the plane locally orthogonal to $\Bv$.  The adiabatic invariant
\begin{equation}
		\mu \equiv \frac{mv_{\perp}^2}{2B}\;,
		\label{eq:mudef}
\end{equation}
is conserved to high degree of approximation, providing an approximate 
third integral of the motion.  The unperturbed dynamics of this class 
of orbit is thus quasi-integrable, not chaotic, implying that we must 
use quasilinear theory of type QLT2 to derive a diffusion equation 
(i.e. heating will occur only beyond a nonlinear amplitude threshold).  
Cyclotron resonance heating in mirror geometries has been much 
discussed in the literature \cite[pp.  
413--422]{lieberman-lichtenberg94} and we shall not discuss the 
heating of the trapped particles in this paper.

\subsection{Free particles}
\label{sec:Freeorb}

For a given energy $\calE$ and conserved momentum $p_z$, the 
\emph{energetically accessible region} is the set of points $(x,y)$ 
for which there exist $p_x$ and $p_y$ such that $H_0(x,y,p_x,p_y,p_z) 
= \calE$.  From \eq{eq:H0} we get the inequality
\begin{equation}
		[p_z - q\psi(x,y)]^2 \leq 2m\calE \;.
		\label{eq:Eaccess}
\end{equation}

We define the free particles as those for which the origin is 
energetically accessible.  Thus, since $\psi(0,0)=0$, \eq{eq:Eaccess} 
implies that free particles are those for which
\begin{equation}
		p_z^2 \leq 2m\calE \;.
		\label{eq:freep_z}
\end{equation}

The total region accessible to free particles is the set of $(x,y)$ 
for which the ranges of $p_z$ defined by \eq{eq:Eaccess} and 
\eq{eq:freep_z} are not disjoint.  This gives the condition
\begin{equation}
		|q\psi(x,y)| \leq 2(2m\calE)^{1/2} \;.
		\label{eq:totalfree}
\end{equation}
This inequality being satisfied, the intersection of the ranges 
defined by \eq{eq:Eaccess} and \eq{eq:freep_z} is 
$[q\psi-(2m\calE)^{1/2},(2m\calE)^{1/2}]$ for $q\psi>0$, 
or $[-(2m\calE)^{1/2},q\psi+(2m\calE)^{1/2}]$ for 
$q\psi<0$.

\begin{figure}[tbp]
		\centering
		\includegraphics[scale=0.5]{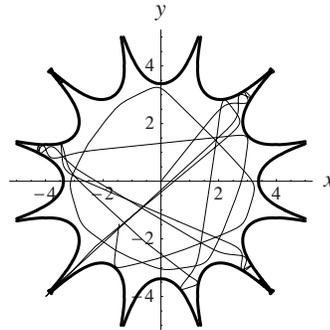}
		\caption{A typical free-particle orbit in the magnetic field shown in 
		\protect\fig{fig:Contours}, with initial conditions as 
		described in the text.}
		\label{fig:ChaoticOrbit}
\end{figure}

A typical trajectory for a free particle with energy well below the 
escape energy is shown in \fig{fig:ChaoticOrbit}.  In this case $p_z = 
0$, and the initial conditions are $x_0 = y_0 = 0$, $p_{x0} = 0.04$, 
$p_{y0} = 0.05$, giving $\calE = 0.00205$.

We see that the energetically accessible region is bounded by a series 
of curves (shown as thick lines) joining in cusps at the magnets.  
These bounding curves are convex toward the confinement region, except 
at the cusps, where the magnetic field goes to infinity so that no 
particle can penetrate.  Also we see that the motion of the particle 
well away from the bounding curves is approximately rectilinear so 
that the system does indeed appear like a physical realization of a 
Sinai billiard system \cite{sinai70}.

\begin{figure}[tbp]
		\centering
		\includegraphics[scale=0.5]{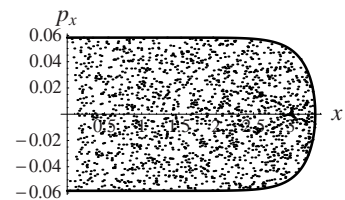}
		\caption{Intersections of an orbit with the Poincar\'{e} surface 
		of section $y=0$ described in the text.}
		\label{fig:ChaoticSea}
\end{figure}

Although the orbit in \fig{fig:ChaoticOrbit} looks chaotic, a better 
test for chaos is to do a Poincar\'{e} surface of section puncture 
plot, as shown in \fig{fig:ChaoticSea} for a particle with $p_z = 0$ 
and energy $\calE = 0.0017$ started near the periodic orbit 
described in Sec.~\ref{sec:Perorb}.  The Poincar\'{e} surface of 
section is $x>0$, $y=0$ and its images under the symmetry operation 
$\zeta \mapsto \exp(\ij \pi/6)\zeta$.  Dots indicate both upward- and 
downward-going passes of the orbit.  It is seen that the orbit appears 
to fill the energetically accessible phase space ergodically, 
indicating strong chaos.

However, the interaction with the field near the bounding curve is not 
specular reflection with a zero radius of curvature, so our magnetic 
cusp confinement system is not precisely analogous to that studied by 
Sinai.  In fact, studying \fig{fig:ChaoticOrbit} we see that there are 
two qualitatively different kinds of reflection event --- 
approximately specular reflection with a small-but-finite radius of 
curvature, and ``sticky reflections'' near the cusps, where the 
particle is temporarily trapped in a one-sided mirror field and 
oscillates several times before reflection.  As explained below, we 
call these nonadiabatic and adiabatic reflections, respectively.

\subsubsection{Scattering analysis of reflection, large $n$}
\label{sec:Scattering}

In order to make the reflection process a precisely defined event we 
go to the large-$n$ limit, in which the spacing between the magnets, 
$\pi a/n$, and the scale length of magnetic field variation, $a/n$, 
become small compared with the radius $a$.  Since we are interested in 
dynamics near the wall of magnets, we shift the origin to the saddle 
point at $\zeta = a$ by defining $\xi \equiv \zeta - a$, just as in 
Sec.~\ref{sec:Nearfield}.  Again \eq{eq:PFapprox} applies at 
leading order, but this time its region of validity extends beyond the 
region of the magnetic null to include the high-field regions near the 
magnets (the only requirement being $|\xi| \ll a$).

We now take \eq{eq:PFapprox} to be the \emph{exact} complex flux 
function for the \emph{model problem} of an infinite line of magnets 
(treating $n$ as an arbitrary parameter, which is scaled out in 
the nondimensionalization defined in Sec.~\ref{sec:Effpot}).  A 
\emph{reflection event} is now precisely defined as a scattering 
process, in which a particle impinges from $\re\xi = -\infty$ (where 
the orbit is asymptotically a straight line), reflects off the 
magnetic field of the magnets, then retreats back toward $\re\xi = 
-\infty$.

\begin{figure}[tbp]
		\centering
		\includegraphics[scale=0.5]{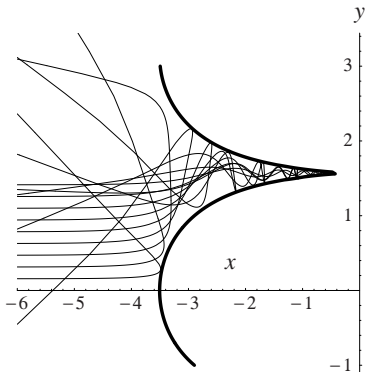}
		\caption{A set of nine orbits, incident in the direction normal to 
		the line of magnets, with different ``impact parameter'' $y_0$.  
		The thick line shows the boundary of the energetically accessible 
		region. The origin of the $x$-axis has been shifted to lie on the 
		line of magnets.}
		\label{fig:Scattering}
\end{figure}
In \fig{fig:Scattering} we illustrate some typical reflection events 
for particles with initial momentum $p_{0x} = 0.06$, $p_{y0} = 0$, 
$p_z = 0$, giving energy $\calE = 0.0018$.  For small initial 
values of $y$, $y_0$, the reflection is approximately specular, but as 
$y_0$ approaches $\pi/2$ (the height of the first magnet above the 
$x$-axis), the orbit undergoes more and more oscillations (gyrations) 
before being reflected back.

Clearly, for $y_0 \approx \pi/2$ we can use adiabatic invariant theory 
(cf.  Sec.~\ref{sec:Traporb}) to treat the process of reflection in 
the mirror field in the throat of the cusp field, whereas for $y_0 
\approx 0$ the particle does not complete even one gyration about the 
magnetic field, so the adiabatic invariant is not defined on any part 
of the orbit.  In order to determine on which part of any given orbit 
$\mu$ is approximately conserved, we compute the cyclotron frequency 
$f_{\mathrm{c}} \equiv \omega_{\mathrm{c}}/2\pi$ at each point on the 
orbit, where $\omega_{\mathrm{c}}(t) \equiv |q|B(t)/m$, and compare it 
with the time-rate-of-change of $\ln B$.

Suppose that the inequality $f_{\mathrm{c}} > \dot{B}/B$ holds over 
the interval $t_1 < t < t_2$ and is violated outside the interval.  
Adiabatic invariance theory applies during this interval, provided the 
particle has enough time to execute at least one gyroorbit.  To 
determine the latter point, we calculate the total gyrophase change 
over the interval in which adiabatic invariance potentially applies,
\begin{equation}
		\Delta\phi \equiv \int_{t_1}^{t_2}\omega_{\mathrm{c}}\,\d t \;.
		\label{eq:Delphidef}
\end{equation}
Then we define an \emph{adiabatic reflection} as one for which 
$\Delta\phi/2\pi > 1$ and a \emph{nonadiabatic reflection} as one for 
which $\Delta\phi/2\pi \leq 1$.

\begin{figure}[tbp]
		\centering
		\includegraphics[scale=0.5]{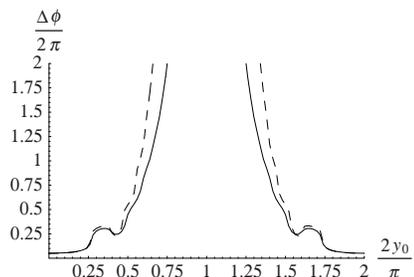}
		\caption{Adiabaticity parameter $\Delta\phi/2\pi$ \emph{vs.} 
		impact parameter times $2/\pi$ for normally incident orbits with 
		energy $0.01$ (solid line) and very low energies (dashed line).}
		\label{fig:Horizontal}
\end{figure}
Figure \ref{fig:Horizontal} shows this adiabaticity parameter for the 
case of normal incidence, as depicted in \fig{fig:Scattering}. Two 
values of energy are shown, a relatively high energy $\calE = 
0.01$ and the low-energy limit $\calE \rightarrow 0$ (see below). 
Reflection is nonadiabatic for roughly $60\%$ of particles in 
both cases.

\begin{figure}[tbp]
		\centering
		\includegraphics[scale=0.5]{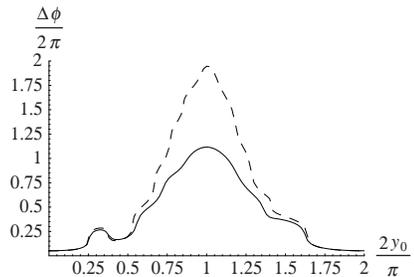}
		\caption{Adiabaticity parameter $\Delta\phi/2\pi$ \emph{vs.} 
		impact parameter times $2/\pi$ for orbits incident at $20^{\circ}$ 
		to the normal with energy $0.01$ (solid line) and very low 
		energies (dashed line).}
		\label{fig:Oblique}
\end{figure}
Figure \ref{fig:Oblique} shows the dependence of $\Delta\phi/2\pi$ on 
impact parameter $y_0$ for particles incident at an angle of 
$20^{\circ}$ to the normal in the $x$-$y$ plane, and with $p_z = 0$.  
(Here $y_0$ is defined such that the orbits have initial values $x = 
X_0$, $y = Y_0 + y_0$, where $X_0$ is an arbitrarily large negative 
constant and the constant $Y_0$ is chosen so that $y_0 = 0$ 
corresponds to an orbit symmetric about the $x$-axis.)

It is seen that the adiabatic region is much reduced in the low-energy 
case, and has virtually disappeared in the high-energy case.  At 
angles of incidence greater than $25^{\circ}$, both high- and 
low-energy particles reflect nonadiabatically for all impact 
parameters.  The ratio of the solid angle occupied by the cone of 
angles of incidence $\eta < 25^{\circ}$ to the cone of all possible 
angles of incidence, $\eta < 90^{\circ}$ is about $0.093$.  Thus we 
conclude that considerably less than $10\%$ of particles reflect 
adiabatically.

The low-energy limit referred to above is defined by observing that 
the lower the incident energy, the larger the value of $-\re \xi$ at 
which the particle reflects.  Thus, in this limit we can replace 
$\sech(n\xi/a)$ by $2\exp(n\xi/a)$ in \eq{eq:PFapprox} and define the 
low-energy approximation as the result of replacing $\Psi$ with
  \begin{eqnarray}
    \Psi_{\mathrm{low}} \equiv 2\psi_{\mathrm{X}}\,\exp\!
    \left(
      \frac{n\xi}{a}
    \right) \;,
    \label{eq:PFlow}
  \end{eqnarray}
where $\psi_{\mathrm{X}}$ is defined in \eq{eq:psiXdef}.

The dynamics in this limit exhibits an important scale invariance: if 
we displace the orbit in the $x$-direction using the transformation 
$\xi = \xi' + h$, where $h$ is a real constant, then the flux function 
changes by a constant factor: $\Psi_{\mathrm{low}}(\xi) = 
\exp(nh/a)\Psi_{\mathrm{low}}(\xi')$.  Inspecting \eq{eq:complexHeqns} 
we see that the transformation $\pv = \exp(nh/a)\pv'$, $t = 
\exp(-nh/a)t'$ leaves the form of the equations of motion invariant.  
The energy is transformed according to $\calE = 
\exp(2nh/a)\calE'$.  Clearly $y_0$ is invariant and it is also 
easy to show from \eq{eq:Delphidef} that $\Delta\phi$ is invariant 
under this scaling transformation.  Thus we have the result that 
\emph{in the low-energy limit} the function $\Delta\phi(y_0)$ \emph{is 
independent of energy}.

\subsubsection{$z$-Motion}
\label{sec:zMotion}

\begin{figure}[tbp]
		\centering
		\includegraphics[scale=0.5]{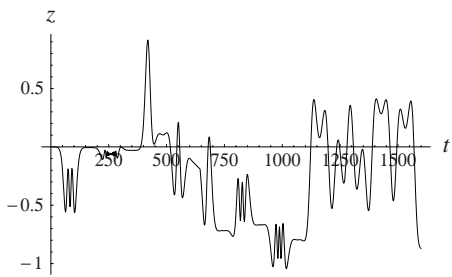}
		\caption{The $z$-component of the orbit shown in 
		\protect\fig{fig:ChaoticOrbit}.}
		\label{fig:ChaoticOrbitz}
\end{figure}

Although our idealized system is infinitely long, any real system will 
be of finite length and it is therefore of interest to enquire as to 
the rate of drift in the $z$-direction.  
Figure~\ref{fig:ChaoticOrbitz} shows the motion in $z$ for the case 
shown above.  We see that, for the case $p_z = 0$, the motion appears 
to be a random walk with no secular drift.

\subsubsection{Periodic orbits}
\label{sec:Perorb}

As for the four-cusp system of Dahlqvist and Russberg 
\cite{dahlqvist-russberg90}, we show that our $n=6$ example is not 
\emph{completely} chaotic by showing numerically that there is at 
least one stable orbit surrounded by invariant tori 
(Kolmogorov--Arnol'd--Moser or KAM surfaces) which make a small region 
around the periodic orbit dynamically inaccessible to the chaotic 
orbit filling most of the energy surface.

\begin{figure}[tbp]
		\centering
		\includegraphics[scale=0.5]{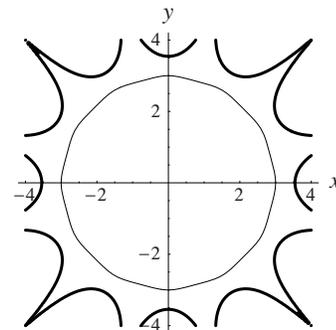}
		\caption{Stable periodic orbit with 12-fold symmetry --- the 
		``dodecagonal'' orbit --- as described in the text.  The thick 
		line shows the boundary of the energetically accessible region.}
		\label{fig:12agonalOrbit}
\end{figure}

We expect the most stable orbit to be the one with the highest 
symmetry allowed by the system, i.e. $2n$-fold symmetry, since this is 
the smoothest orbit, least like the trajectory of a billiard ball.  
This is illustrated in \fig{fig:12agonalOrbit} for the case $n=6$, 
$p_z=0$, with the orbit passing through $(x,y)_0=(3,0)$.  The 
corresponding momentum required to close the orbit is $(p_x,p_y)_0 = 
(0,0.04925)$, giving an energy $\calE = 0.00170082$ and a period 
$28.96$.  Because of its 12-fold symmetry we call this the dodecagonal 
orbit.

To investigate the stability of such an orbit we linearize about the 
orbit and calculate the eigenvalues of the matrix evolving a 
neighborhood of the phase-space point $(x_0,p_{y0})$ in the $x>0$, 
$y=0$ half plane to its intersection with the next surface of section 
that is equivalent under the symmetry operation $\zeta \mapsto 
\exp(\ij \pi/6)\zeta$.  [The crossing time is found numerically by 
searching for the first zero of $\arg\exp(-\ij \pi/6)\zeta(t)$.]  For the 
case shown in \fig{fig:12agonalOrbit}, the eigenvalues are $-0.54162 
\pm 0.840624 \ij$, which lie on the unit circle in the complex plane, 
indicating stability.

\begin{figure}[tbp]
		\centering
		\includegraphics[scale=0.5]{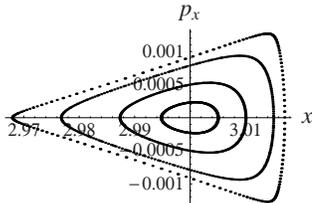}
		\caption{Poincar\'{e} section transverse to the dodecagonal orbit 
		showing several satellite orbits which appear to be on invariant 
		tori.}
		\label{fig:12agonalOrbitPSec}
\end{figure}

This stability is confirmed more graphically by the Poincar\'{e} plots 
in \fig{fig:12agonalOrbitPSec} for some neighboring orbits on the same 
energy surface as the dodecagonal orbit shown in 
\fig{fig:12agonalOrbit}.  Figure~\ref{fig:12agonalOrbitPSec} shows 
iterates of the map $(x,p_x) \mapsto (x',p'_x)$ found by calculating 
the crossing time $t$ with the $\theta = \pi/6$ line as described 
above, then calculating $x' = \re\exp(-\ij \pi/6)\zeta(t)$, $p'_x = 
\re\exp(-\ij \pi/6)p_{\zeta}(t)$.

Figure~\ref{fig:12agonalOrbitPSec} shows an island of regular 
motion in a vast chaotic sea --- if we start much beyond the last 
orbit shown, the orbit rapidly moves far away from the periodic orbit.  
For example, the chaotic orbit shown in the Poincar\'{e} plot, 
\fig{fig:ChaoticSea}, started at $(x,y)_0=(3.02,0)$, with $p_x = 0$ and 
$p_y$ adjusted to give the same energy as that of the periodic orbit 
plotted in \fig{fig:12agonalOrbit}.

The orbits in \fig{fig:12agonalOrbitPSec} appear to lie on invariant 
curves that are topologically circular, being the intersection of 
invariant tori with the surface of section.  The quasi-triangular 
shape of the outer orbits is due to the existence of three unstable 
periodic X-points which define the separatrix between regular and 
chaotic motion (cf.  the bifurcation with branching ratio 1/3 in 
Fig.~1(c) of \cite{greene-etal81}).

Scanning through a range of initial values of $x$, so as to vary the 
energy $\calE$, we find that for all $\calE$ less than the 
escape energy $\calE_{\mathrm{esc}}$ (and somewhat beyond) 
the dodecagonal orbit is stable.  We have not done an exhaustive study 
of other periodic orbits, but have established that the 
4-fold-symmetric ``square'' orbit is unstable below an energy 
of about 0.1, but stabilizes above this value.

The existence of a stable periodic orbit shows that the multiple-cusp 
confinement system is \emph{not} completely chaotic.  However, the 
islands around the few stable orbits are very small.  For instance the 
area occupied in the $x-p_x$ plane by the island shown in 
\fig{fig:12agonalOrbitPSec} is about $7 \times 10^{-5}$, compared with 
the energetically accessible area of the Poincar\'{e} section, $\int 
[2m\calE-q^2\psi^2(x,0)]^{1/2})\,\d x \approx 0.37$, i.e. four orders 
of magnitude smaller.  Even a very small amount of extrinsic 
stochasticity (e.g. small-angle collisions with other particles) will 
easily destroy such small islands of stability.

Thus we conclude that, for practical purposes, the assumption of 
complete chaos in the free-particle dynamics is well justified, and 
hence assume that any orbit covers its energy surface ergodically.

\section{Wave-particle interaction model}
\label{sec:Wave}

For a given wavelength, $\lambda$, of the incident wave, the ratio 
$\lambda/(a/n)$ tends to infinity as $n \rightarrow \infty$.  Hence we 
consider the long-wavelength limit, in which the wave-particle 
interaction is via the uniform, oscillatory electric field $\Ev = 
\re(\Eosc\exp\ij\omega t)$, where $\Eosc$ is a constant complex 
vector.  On the other hand we assume $\omega \gg 
\omega_{\mathrm{p,e}}$, where $ \omega_{\mathrm{p,e}}$ is the electron 
plasma frequency, so that the oscillatory part of the electrostatic 
potential can be ignored.  Thus the electric field is taken to be 
produced by the vector potential $\Av_{1} = 
\re[\ij(\Eosc/\omega)\exp\ij\omega t]$.

In reality, uniformity of $\Ev$ applies only during one 
wall-scattering event, which occurs over the scale length $a/n$, 
whereas $\Ev$ can be different at different points on the 
picket fence if $\lambda$ is comparable with $a$.  In this 
paper we consider a model system in which $\Ev$ is strictly 
constant in space, but we model the real situation by choosing $
\Eosc$ from a random ensemble of values with probability 
distribution matching the distribution of field amplitudes and phases 
actually encountered on the cylinder $r = a$.  We also assume the 
distribution of initial phases to be uniform, which means that the 
\emph{ensemble averaging operator} $\langle\cdot\rangle$ 
\emph{automatically includes phase averaging}.

The model Hamiltonian determining the full particle dynamics is thus 
taken to be $H_0 + H_1$, where $H_0$ is given 
by \eq{eq:H0}, and
  \begin{eqnarray}
    H_1 \equiv -\frac{q}{m\omega}\re[\ij(\pv - 
    q\psi\ev_{z})\cdot\Eosc\exp \ij\omega t] \;,
    \label{eq:H1}
  \end{eqnarray}
with $\ev_{z}$ being the unit vector in the $z$-direction.
The above form for $H$ follows simply by expanding 
$(\pv-q\Av)^{2}/2m$ and dropping the term quadratic in 
$\Eosc$ as it has no spatial dependence and thus does not affect 
the dynamics. (If we had not taken $\Ev$ to be constant in 
space, it would be necessary to retain this term to include
ponderomotive force effects.)

Note that $H_1$ produces an oscillatory correction, $\partial 
H_1/\partial \pv$, to the velocity, $\dot{\xv}$, that is independent 
of position, whereas the oscillation in $\dot{\pv}$ is localized 
around the edge region due to the localization of $\psi$.  Thus $\pv$ 
is nonoscillatory in the middle region where $\psi$ is negligible.  
This is a consequence of our choice of gauge for representing $\Ev$, 
which automatically gives an oscillation-center representation 
\cite[p.  47]{schmidt66} in generalized momentum space.  The 
localization of $\dot{\pv}$ to the edge region where the particles are 
reflected is desirable since this is the region where irreversibility 
is introduced --- in the middle region the particles simply respond 
adiabatically to the high-frequency field.

It is possible also to remove the oscillation in the generalized 
position coordinate in the middle region, so as to completely localize 
the interaction to the edge region, by making a canonical 
transformation to oscillation-center variables \cite{dewar73}.  
However, as we are interested in the diffusion in energy, not 
position, we have not found this transformation to be useful.

\section{Quasilinear diffusion}
\label{sec:Transport}

The Vlasov equation for the single-particle distribution function
$f(\xv,\pv,t)$ can be written
  \begin{eqnarray}
       \partial_{t}f + \{f,H\} = 0 \;,
    \label{eq:Vlasov}
  \end{eqnarray}
where the Poisson bracket $\{f,H\} \equiv
\partial_{\xv}f\cdot\partial_{\pv}H -
\partial_{\pv}f\cdot\partial_{\xv}H$.

We decompose $f$ into a nonfluctuating part, $\overline{f}$, and a 
fluctuating remainder $\widetilde{f}$.  We define $\bar{f} \equiv 
\langle\langle f\rangle\rangle$ where 
$\langle\langle\cdot\rangle\rangle$ includes not only an average over 
the wave phase via ensemble averaging, but a phase-space average over 
the unperturbed energy surface: for arbitrary phase-space function 
$g(\xv,\pv,t)$, varying on both the fast timescale 
$\omega^{-1}$ and a slower (diffusion) timescale, we define 
$\langle\langle g\rangle\rangle(\calE,p_z,t)$, varying only on 
the slow timescale, by
  \begin{eqnarray}
       \langle\langle g\rangle\rangle
       \equiv
          \frac{1}{\calN}
          \int_{r<a}\!\!\!\d^{2}\xv\int\!\!\d^{2}\pv\;
          \delta(\calE - H_0(\xv,\pv))
          \langle g(\xv,\pv,t)\rangle \;,
    \label{eq:Proj}
  \end{eqnarray}
where $r \equiv ({x}^{2}+{y}^{2})^{1/2}$, $\d^{2}\xv
\equiv \d x \d y$, $\d^{2}\pv \equiv \d p_x\d p_y$,
$\delta$ denotes the 
Dirac $\delta$-function, and the normalizing factor 
$\calN(\calE,p_z)$ is defined by
  \begin{eqnarray}
       \calN \equiv 
       \int_{r<a}\!\!\!\d^{2}\xv\int\!\!\d^{2}\pv\;
       \delta(\calE - H_0(\xv,\pv)) \;.
    \label{eq:Norm}
  \end{eqnarray}
  
The result of applying this projection operation is a function only of 
$\calE$ and $t$, so that $\{\overline{f},H_0 \} \equiv 0$ and 
hence the averaged part of the distribution function, $\overline{f}$, 
is invariant under the unperturbed dynamics.  The projection of the 
distribution function onto the energy surface using the averaging 
operator is an extreme form of phase-space coarse-graining.  Owing to 
the highly chaotic nature of the unperturbed dynamics we assume that 
this coarse-grained distribution function relaxes to a function of the 
constants of the unperturbed motion, $\calE$ and $p_z$, on a 
timescale much faster than the quasilinear diffusion timescale.  We 
assume all particles to be confined, so that the region of phase space 
defined by $\calE = \const$, $p_z = \const$ is bounded within $r 
< a$.
 
Applying the operation $\langle\langle\cdot\rangle\rangle$ to 
\eq{eq:Vlasov} we have
  \begin{eqnarray}
       \partial_{t}\overline{f}
     + \langle\langle\{ \widetilde{f},H_1 \}\rangle\rangle
     = 0 \;.
    \label{eq:VlasovAv}
  \end{eqnarray}

Writing the Poisson bracket in the form
 $\{\widetilde{f},H_1 \} \equiv
\partial_{\xv}\cdot(\widetilde{f}\partial_{\pv}H_1) -
\partial_{\pv}\cdot(\widetilde{f}\partial_{\xv}H_1)$
and integrating by parts (assuming the particle confinement is good 
enough that the boundary contribution can be ignored) we find
  \begin{eqnarray}
       \langle\langle\{ \widetilde{f},H_1 \}\rangle\rangle & = &
       \frac{1}{\calN}
       \frac{\partial}{\partial \calE}
       \left(
          \calN
          \langle\langle\,
            \widetilde{f}\dot{\calE}
          \rangle\rangle
       \right) \nonumber \\
			 & & \mbox{} +
       \frac{1}{\calN}
       \frac{\partial}{\partial p_z}
       \left(
          \calN
          \langle\langle\,
            \widetilde{f}\dot{p}_z
          \rangle\rangle
       \right)
        \;.
    \label{eq:PBAv}
  \end{eqnarray}
where $\dot{\calE}$ is the rate of change in the energy integral 
of the unperturbed system, $\calE(t) \equiv 
H_0(\xv(t),\pv(t))$, along the perturbed orbit.  Noting that 
$\dot{H_0} = \{ H_0, H_0 + H_1 \} \equiv \{ H_0,H_1 \}$, we see that
	\begin{equation}
			\dot{\calE} = \{H_0,H_1 \} \;.
			\label{eq:Edot}
	\end{equation}
Also, $\dot{p}_z \equiv \{p_z, H_0+H_1 \} = \{p_z, H_1 \}$ [which 
vanishes for our simple interaction term, \eq{eq:H1}].

Subtracting \eq{eq:VlasovAv} from \eq{eq:Vlasov} we also have
  \begin{eqnarray}
       \partial_{t}\widetilde{f} + \{ \widetilde{f},H_0 \}
       = -\{ \overline{f},H_1 \} + O(\Eosc^{2}) \;.
    \label{eq:VlasovOsc}
  \end{eqnarray}
Linearizing \eq{eq:VlasovOsc} and solving by integration along the 
unperturbed trajectories from an initial time $-T$, we have
  \begin{eqnarray}
       \widetilde{f}(\xv,\pv,t) & = & \widetilde{f}(\xv(-T),\pv(-T),-T)
			 \nonumber \\ & & \mbox{}
       -\int_{-T}^{t}\d t'\;
			 \left[
	       \dot{\calE}'
				 \frac{\partial{\overline{f}}}{\partial{\calE}}(\calE',p'_z,t')
			 \right.
			 \nonumber \\ & & \mbox{}
			 \left.
				 +\dot{p}'_z
				 \frac{\partial{\overline{f}}}{\partial{p_z}}(\calE',p'_z,t')
			 \right]
       \;.
    \label{eq:fOsc}
  \end{eqnarray}
In calculating $\dot{\calE}' \equiv \dot{\calE}(t')$ using 
\eq{eq:Edot}, the right hand side is to be evaluated at the point 
$(\xv(t'),\pv(t'))$ on the unperturbed phase-space trajectory that 
passes through $(\xv,\pv)$ at time $t$, and similarly for 
$\dot{p}_z(t')$ if an interaction model is used for which it is 
nonvanishing.

We now observe that, for large $T$, 
$\widetilde{f}(\xv',\pv',t')(t'=-T)$ becomes decorrelated from 
$\dot{\calE}(t)$ and $\dot{p}_z(t)$ and thus does not contribute to 
the averages on the right hand side of \eq{eq:PBAv}.  The 
decorrelation time is the duration of one wall-scattering event, which 
is of the order of the \emph{transit time}
\begin{equation}
		\tau_{\mathrm{tr}}(\calE)
		\equiv 
		\frac{a}{n|v|}
		\label{eq:transitTime}
\end{equation}
of a free particle with speed $|v| \equiv (2\calE/m)^{1/2}$ through 
the scale length $a/n$ of the magnetic field variation.  Thus, 
assuming $T \gg \tau_{\mathrm{tr}}$, we can set the initial value term 
$\widetilde{f}(\xv',\pv',t')(t'=-T)$ to zero without significant error.

Also, if $\xv(t)$ is in the wall-interaction region, where 
$\dot{\calE}$ and $\dot{p}_z$ are significant, then $\xv(-T)$ is far 
from the wall so $\dot{\calE}(-T)$ and $\dot{p}_z(-T)$ are negligible 
(because $\psi$ is essentially zero there --- see the discussion at 
the end of Sec.~\ref{sec:Wave}).  Thus we can, to a very good 
approximation, extend the lower limit of the integral in \eq{eq:fOsc} 
to $-\infty$.

Whereas $T$ is assumed large with respect to $\tau_{\mathrm{tr}}$, we 
assume it to be small with respect to the characteristic evolution 
time for the distribution function $\overline{f}$.  (That is, we 
assume the wave to be of sufficiently low amplitude that it takes many 
wall-interaction events for significant heating to occur.)  Thus we 
can also make the \emph{Markovian approximation} that 
$\overline{f}(\calE',p'_z,t')$ can be moved outside the integral in 
\eq{eq:fOsc} with negligible error.

Substituting \eq{eq:fOsc} in \eq{eq:PBAv} and then \eq{eq:VlasovAv} we 
find (assuming $\dot{p}_z = 0$) the \emph{quasilinear diffusion 
equation}
  \begin{eqnarray}
       \frac{\partial\overline{f}}{\partial t}
     = \frac{1}{\calN}\frac{\partial}{\partial\calE}
     \left(
      \calN D \frac{\partial\overline{f}}{\partial \calE}
     \right) \;,
    \label{eq:QLDeq}
  \end{eqnarray}
where $D(\calE,p_z)$ is the energy diffusion coefficient,
defined by
  \begin{eqnarray}
       D \equiv \half\int_{-\infty}^{\infty} C(\tau) \d\tau \;,
    \label{eq:QLDcoeff}
  \end{eqnarray}
with the two-time correlation function $C(\calE,p_z,\tau)$
  \begin{eqnarray}
       C(\tau) \equiv
       \langle\langle
          \dot{\calE}(t-\tau)
          \dot{\calE}(t)
       \rangle\rangle
			 =
       \langle\langle
          \dot{\calE}(\tau)
          \dot{\calE}(0)
       \rangle\rangle \;,
    \label{eq:CorrelFn}
  \end{eqnarray}
where the second form follows from the fact that, because of the 
average stationarity of the dynamical system, $C$ depends only on the 
time difference, $\tau = t - t'$.  Also note that the projection 
operation $\langle\langle\cdot \rangle\rangle$, \eq{eq:Proj}, can be 
done using either initial or final values as independent variables 
because the Jacobian of the transformation is unity (preservation of 
phase space volume) and $H_{0}$ is an invariant of the unperturbed 
motion.  Thus $C(\tau)$ is an even function, which fact we used to 
extend the upper limit of the integral in \eq{eq:QLDcoeff} to 
infinity.  We can also use time reversal invariance to show 
$C(\calE,p_z,\tau) = C(\calE,-p_z,-\tau) = 
C(\calE,-p_z,\tau)$.

We end this section by calculating the heating rate due to Fermi 
acceleration. First we define the total plasma energy per unit length
in the $z$-direction,
  \begin{eqnarray}
       U(t) \equiv 
        \int_{0}^{\infty}\d \calE
        \int_{-\infty}^{\infty}\d p_z\;
        \calN\calE\overline{f} \;.
    \label{eq:TotEn}
  \end{eqnarray}
Differentiating $U$ with respect to time, using \eq{eq:QLDeq} and 
integration by parts we find 
the rate of power deposition into the plasma due to reflections from 
the confining edge magnetic field under the influence of an 
electromagnetic wave
  \begin{eqnarray}
       \dot{U}(t) = 
        -\int_{0}^{\infty}\d \calE
        \int_{-\infty}^{\infty}\d p_z
        \calN D
        \frac{\partial\overline{f}}{\partial\calE}\;.
    \label{eq:Heating}
  \end{eqnarray}
We have assumed that $\partial\overline{f}/\partial\calE$ 
vanishes at an energy less than or equal to the maximum confined 
energy $q^{2}\psi_{\mathrm{X}}^{2}/2m$ discussed in 
Sec.~\ref{sec:Nearfield} so that we can ignore boundary contributions.

\section{One-dimensional model}
\label{sec:Simple}

We saw in Sec.~\ref{sec:Scattering} that most particles 
reflect nonadiabatically in less than one gyroperiod, and thus should 
not be sensitive to the details of the $y$-variation of $\psi$ (i.e. 
subtle resonance effects should not be important for most particles).  
This suggests we estimate the effect of nonadiabatic reflection by 
using a one-dimensional model Hamiltonian obtained by replacing 
$\psi(r,\theta)$ in \eq{eq:H0} with an axisymmetric flux function, 
$\psi(r)$ (cf.  the one-dimensional model used by Yoshida \emph{et 
al.} \cite{yoshida-etal98}). The conservation of the angular momentum 
$p_{\theta}$ then allows formal integration of the equations of 
motion by the method of quadratures.

Although such a one-dimensional flux function violates Laplace's 
equation, and therefore would require a plasma current to produce it, 
this fact is irrelevant to the single-particle dynamics.  By suitable 
choice of $\psi(r)$ we can model the gross radial confinement 
properties of the two-dimensional flux function.  The main loss in the 
physics is that $\psi(r)$ produces no radial component of $\Bv$, and 
hence no interaction with $\widetilde{E}_{\theta}$.  But if we assume 
perfectly conducting wall boundary conditions, $\Eosc$ must be purely 
radial at the vacuum vessel wall (assumed just inside the array of 
magnetic dipoles) so $\widetilde{E}_{\theta}$ (and 
$\widetilde{E}_{z}$) can be assumed to be small in the interaction 
region anyway.

We further simplify the reflection dynamics by going to the 
large-$n$ limit, so that the dipoles become a linear array and we can 
use Cartesian coordinates as in Sec.~\ref{sec:Scattering}.  Also, 
Figs.~\ref{fig:Horizontal} and \ref{fig:Oblique} indicate that the 
low-energy approximation, \eq{eq:PFlow}, is good for nonadiabatic 
reflections.  In this limit the field strength $B$ is independent of 
$y$, so we define the equivalent one-dimensional flux function 
$\psi(x)$ as that which gives the same field strength $B(x)$.  That 
is, $B = |\Psi_{\mathrm{low}}'(\xi)| \equiv \psi'(x)$, where
  \begin{eqnarray}
    \psi(x) = 2\psi_{\mathrm{X}}\,\exp\!
    \left(
      \frac{nx}{a}
    \right) \;.
    \label{eq:PFlow1D}
  \end{eqnarray}
(In the above we have shifted the origin of the $x$-axis to lie on the 
same line as the array of dipoles.)  As a final simplification of the 
unperturbed dynamics we evaluate $D$ only at $p_z = 0$.  That is, we 
consider only unperturbed orbits having the constant of the motion 
$p_z = 0$.

In the large-$n$ limit the boundary region where $\psi$ is not small 
makes negligible contribution to \eq{eq:Norm} and thus we find the 
normalizing factor $\calN$ to be independent of energy,
  \begin{eqnarray}
     \calN
     = 2\pi^2 a^2 m \;.
  \label{eq:Napprox}
  \end{eqnarray}

The equations of motion \eq{eq:complexHeqns} can be integrated 
explicitly to give
\begin{eqnarray}
		x(t) & = & \left(\frac{a}{n}\right) \ln
		\left(
		u\,\sech 
			\frac{2u(t - t_{\mathrm{max}})}{\tau_{\mathrm{X}}}
		\right)
		\label{eq:1Dx}  \\
		p_x(t) & = & -2u\left(\frac{m a}{n\tau_{\mathrm{X}}}\right)
		\tanh 
			\frac{2u(t - t_{\mathrm{max}})}{\tau_{\mathrm{X}}} 
		\label{eq:1Dp}
\end{eqnarray}
where $\tau_{\mathrm{X}}$ is defined in \eq{eq:tUnit} and the 
constants of integration are $u \equiv \exp(n x_{\mathrm{max}}/a)$ and 
$t_{\mathrm{max}}$.  Inspecting \eq{eq:1Dx} we see that 
$x_{\mathrm{max}}$ is the maximum value of $x$ attained over the 
entire orbit, and $t_{\mathrm{max}}$ is the time at which this point 
is reached.

Assuming a perfectly conducting vacuum vessel we set $\widetilde{E}_y 
= \widetilde{E}_z = 0$.  Then, from \eq{eq:H1}, $H_1 = 
-(q/m\omega)\re(\ij p_x\widetilde{E}_x\exp \ij\omega t)$ and we have 
the simple expression for the instantaneous power transfer to a 
particle, \eq{eq:Edot},
  \begin{eqnarray}
    \dot{\calE}
    =
    -\dot{p}_x\frac{\partial H_1}{\partial p_x}
    =
    \frac{q}{\omega}\re
    \left[
			\ij\widetilde{E}_x \ddot{x}(t)\exp\ij\omega t
		\right] \;.
    \label{eq:1Dpower}
  \end{eqnarray}

In evaluating the diffusion coefficient using \eq{eq:QLDcoeff}, it is 
convenient first to commute the time integration with the averaging 
operation, so that we first consider the time integral of 
$\dot{\calE}$, which gives the total energy change, $\Delta\calE$, in 
one collision with the wall.  Inserting the analytical solution 
\eq{eq:1Dx} in \eq{eq:1Dpower} and integrating from $t=-\infty$ to 
$+\infty$, we get
  \begin{eqnarray}
    \Delta\calE
    & = &
    -\pi\left(\frac{a}{n}\right)
		{\mathrm{cosech}}\left(\frac{\pi\omega\tau_{\mathrm{X}}}{4u}\right)
		\nonumber \\  && \mbox{} \times
		\re \left(\ij q\widetilde{E}_x\exp\ij\omega t_{\mathrm{max}} \right)
		\;.
    \label{eq:DeltaH}
  \end{eqnarray}

Since \eq{eq:1Dp} expresses the orbit in terms of the constants of 
integration rather than the initial conditions, to evaluate the 
phase-space average we change variables from the initial conditions 
$x,p_x$ to $u$ and $s \equiv u t_{\mathrm{max}}/\tau_{\mathrm{X}}$ so 
$x = (a/n)\ln(u\,\sech 2s)$, $p_x = 2u(ma/n\tau_{\mathrm{X}})\tanh 2s$.
The Jacobian of this transformation is $4(ma^2/n^2\tau_{\mathrm{X}})$, 
so, using \eq{eq:Napprox} in \eq{eq:Proj}, the phase space average 
over wall scattering events is transformed to
\begin{eqnarray}
		\langle\langle \cdot \rangle\rangle
		& = & \frac{8}{\pi}\left(\frac{m a}{n^2\tau_{\mathrm{X}}}\right)
		\nonumber \\  && \mbox{} \times
		 \int_{-\infty}^{\infty}
		 \frac{\d u\;\Theta(\calE-4u^2 \calE_{\mathrm{esc}})}
		 {\sqrt{2m}
		 \left(
		 		\calE-4u^2 \calE_{\mathrm{esc}}
		 \right)^{1/2}}
		 \int_{-\infty}^{\infty}\d s\;
		 \cdot \;,
		\label{eq:phaseint}
\end{eqnarray}
where $\calE_{\mathrm{esc}}$ is defined in \eq{eq:Eescdef} and 
$\Theta(\cdot)$ is the Heaviside step function.

Using \eqs~(\ref{eq:DeltaH}) and (\ref{eq:phaseint}) in 
\eq{eq:QLDcoeff} we have
\begin{equation}
	D = \frac{4}{3\pi}\frac{ q^2\langle|\widetilde{E}_x|^2\rangle}
				{\omega^2}
			\frac{|v|^3}{a}
			G \left(\frac{\pi\omega\tau_{\mathrm{tr}}}{2}
			\right) \;,
	\label{eq:Dxx}
\end{equation}
where $|v| \equiv (2\calE/m)^{1/2}$ is the mean velocity in the 
field-free region and $\tau_{\mathrm{tr}}(\calE)$ is the transit time 
defined in \eq{eq:transitTime}.  (Note that $D$ does not depend on the 
strength of the magnetic field in this model, only the scale length, 
because a change of $\psi_{\mathrm{X}}$ is equivalent simply to a 
shift in the origin of the $x$-axis by an amount of order $a/n$.)

\begin{figure}[tbp]
		\centering
		\includegraphics[scale=0.5]{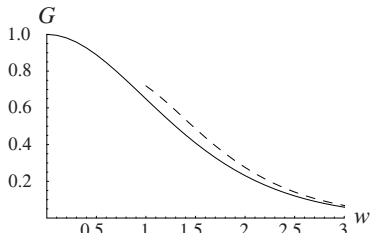}
		\caption{Function $G(w)$ defined in the text (solid line) and 
		the large-$w$ asymptotic approximation (dashed line).}
		\label{fig:G}
\end{figure}
The function $G$ is defined as a one-dimensional integral,
\begin{equation}
		G(w) \equiv 
		\frac{3w^2}{2}\int_{0}^{1}
		\frac{\rho\,{\mathrm{cosech}}^2(w/\rho)}
		{(1-\rho^2)^{1/2}}\d\rho \;,
		\label{eq:Gdef}
\end{equation}
and is plotted as the solid line in \fig{fig:G}.  The function has 
been defined so as to approach unity as $w \rightarrow 0$, as 
discussed below in the context of the low-frequency limit, $\omega \ll 
1/\tau_{\mathrm{tr}}$.  The asymptotic behavior shown by the dashed 
line is discussed below in the context of the high-frequency limit 
$\omega \gg 1/\tau_{\mathrm{tr}}$.

\subsection{Low-frequency (Fermi) limit}

Fermi \cite{fermi49} was concerned with the collision of cosmic rays 
with relatively slowly moving gas clouds.  In our problem this 
corresponds to the low-frequency limit, in which a particle scatters 
off the magnetic field in a time much less than the period of the 
applied field.  This makes the argument of $G$, $w = 
\pi\omega\tau_{\mathrm{tr}}/2$, small.

For $|w| \ll 1$, we can approximate ${\mathrm{cosech}}^2(w/\rho)$ in 
\eq{eq:Gdef} by $\rho^2/w^2$ over nearly the full interval. Evaluating
the integral we find $G(0) = 1$.

Thus, in the low-frequency limit,
\begin{equation}
	D = (4/3\pi) 
	(q^2\langle|\widetilde{E}_x|^2\rangle/\omega^2)(|v|^3/a) \;.
	\label{eq:Dlf}
\end{equation}
This result may be understood as the Fokker-Planck diffusion 
coefficient $\langle(\Delta\calE)^2\rangle/2\tau_{\mathrm{coll}}$ for 
particles oscillating in the applied field with the \emph{quiver 
velocity} $\widetilde{v}$, given by
\begin{equation}
		\widetilde{v} \equiv 
		\frac{q\langle|\widetilde{E}_x|^2\rangle^{1/2}}{m\omega} \;,
		\label{eq:vquiver}
\end{equation}
giving the typical energy step at each collision with the wall 
$\Delta\calE = m|v|\widetilde{v}$.  Taking as a typical time 
between wall collisions $\tau_{\mathrm{coll}} = a/|v|$ we recover, to 
within a factor of order unity, the low-frequency energy diffusion 
coefficient above.

\subsection{High-frequency limit}

In the high-frequency limit, the particle oscillates many times during 
a collision with the magnetic field and we would expect it to respond 
to the applied field adiabatically, gaining little energy.

For $|w| \gg 1$ the dominant contribution to the integral comes 
from a narrow boundary layer near $\rho=1$, in which 
${\mathrm{cosech}}^2(w/\rho)$ may be approximated by 
$\exp(-2|w|)\exp(-2|w|(\rho-1))$. This gives the asymptotic behavior
\begin{equation}
		G(w) \sim
		3\sqrt{\pi}|w|^{3/2}\exp (-2|w|) \;.
  		\label{eq:PhiLargeArg}
\end{equation}

From the dashed line in \fig{fig:G} we see that this asymptotic 
formula gives good agreement with the numerically calculated result 
for $|w|$ greater than about 1.  We see that the energy diffusion is 
indeed exponentially small in this limit.

\section{Discussion}
\label{sec:Discussion}

In this section we give the magnetic parameters of the theory for 
a typical experimental device and make some observations as to the 
possible implications of the theory for such experiments.

Multipolar magnetic cusp confinement has become a conventional method 
for reducing plasma loss on the chamber walls and keeping the inner 
plasma volume free from magnetic field \cite{burke-pelletier92}.  This 
was used in the electron-cyclotron resonance (ECR) plasma formation 
experiment ECRIN (ECR Ions N\'{e}gatifs) 
\cite{ciubotariu-golovanivsky-bacal96,ciubotariu97}.  In ECRIN, 
microwave argon and hydrogen plasmas were excited in a cylindrical 
vessel of internal radius of about 6~cm and length 17~cm surrounded by 
twelve radially magnetized linear permanent magnets of alternating 
polarity, maximum magnetic field strength 0.2~T and microwave cw power 
of 100--1000~W at a frequency of 2.45~Ghz was delivered at one end of 
the vessel.

The primary heating occurred near the microwave input window, but it 
is of interest to consider whether collisionless secondary heating of 
free particles is possible further down the tube, which we can model 
by idealizing the permanent magnets as the $n=6$ linear magnetic 
dipole configuration used for illustration in the present paper in 
Figs.~\ref{fig:Contours}--\ref{fig:ChaoticSea} and 
Figs.~\ref{fig:ChaoticOrbitz}--\ref{fig:12agonalOrbitPSec}.  Using 
$a=6$~cm gives the length unit in these figures (see 
Sec.~\ref{sec:Nearfield}) as $a/n = 1$~cm.  (In this paper we have 
ignored collective effects, collisions and atomic processes, all of 
which may be important in such experiments, so the use of the ECRIN 
parameters should be regarded as illustrative only.)

For ECRIN, the magnetic dipole strength was estimated to be around $K 
= 1.5 \times 10^{-5}\;\mathrm{Tm}^2$.  Using this value in \eq{eq:Eescdef} gives 
the escape energy for electrons as 
$\calE^{\mathrm{e}}_{\mathrm{esc}} \approx 198$~keV, and that 
for singly charged argon ions as 
$\calE^{\mathrm{i}}_{\mathrm{esc}} \approx 2.7$~eV.

For electrons of energy $5$~eV the transit time, \eq{eq:transitTime}, 
is $\tau_{\mathrm{tr}} \approx 7.5$~ns, so that for the microwave 
heating frequency of 2.45~Ghz the argument of the 
transit-time reduction factor $G$ in \eq{eq:Dxx} is $w \equiv 
\pi\omega\tau_{\mathrm{tr}}/2 \approx 182$.  This gives $G(w) \approx 
5.5 \times 10^{-155}$!  Thus we see that nonresonant 
Fermi acceleration is clearly not an important effect in such ECR 
experiments. On the other hand, with $1/\tau_{\mathrm{tr}} \approx 
133$~MHz, this effect can be important in rf heating experiments.

Given the strong transit-time suppression of nonresonant heating, it 
may be of interest to consider the resonant heating of the few free 
electrons penetrating deeply enough into the cusps to reach the ECR 
layer, and this could in principle be calculated using the quasilinear 
formalism developed in this paper.

However we shall content ourselves here simply with estimating the 
proportion of the ECR layer that is accessible to the free electrons, 
as opposed to the trapped electrons discussed in 
Sec.~\ref{sec:Traporb}.  In the neighborhood of the ECR region, 
\eq{eq:totalfree} is satisfied only in the narrow cusp regions directed 
toward the magnets.  We can thus Taylor expand $\psi$ to approximate 
this inequality by $r|\Delta\theta|\omega_{\mathrm{c}}(r) \leq 
2(2m\calE/m)^{1/2}$ in polar coordinates, where 
$\omega_{\mathrm{c}} \equiv |q|B/m$ is the electron cyclotron 
frequency ($=\omega$ in the ECR layer).

Summing these angular ranges over all the $2n$ cusps and dividing by 
$2\pi$ gives the fraction of the ECR layer accessible to free 
particles.  Approximating $r$ by $a$ gives this fraction to be 
$(4/\pi)(\omega\tau_{\mathrm{tr}})^{-1} \approx 1\%$.  On this basis 
we would expect nearly all the ECR power to be deposited in the 
trapped particles, with the free particles being heated through heat 
conduction from the trapped population and other such indirect 
processes.

This paper has focussed only on the effect of chaos as the source of 
stochasticization.  In reality, particle-particle collisions may be 
equally or more important.  Our collisionless energy diffusion 
coefficient will still be valid as an additive contribution to the total 
energy diffusion coefficient provided $\lambda_{\mathrm{mfp}} \gg 
a/n$, for then most particles transit the high-field edge region 
without suffering a collision.  Elastic collisions in the central 
region simply provide a further stochasticization and do not affect 
our result provided the above inequality is satisfied.  Rare 
collisions within the edge region would provide an independent 
additive mechanism for energy diffusion which might or might not 
dominate our collisionless mechanism depending on the ratio of transit 
time to the period of the applied wave.

\section{Conclusion}

We have shown that, in such strongly nonaxisymmetric experiments as 
the multicusp geometry analyzed here, there is a strong collisionless 
stochasticization process due to the chaotic nature of the unperturbed 
particle motion.  This justifies the use of the random phase 
approximation for successive kicks produced by coherent wave-particle 
interaction without having to invoke a nonlinear threshold for 
resonance overlap, or collisions.  Such systems cannot be analyzed by 
area-preserving maps, and thus fall outside the general framework 
usually assumed for the analysis of rf and microwave heating in 
bounded systems \cite{lieberman-godyak98}.

As an alternative to the Fokker-Planck approach for deriving the 
energy diffusion equation we have developed a variant of the 
quasilinear diffusion formalism based on averaging the single-particle 
Liouville equation. This provides a general and efficient formalism 
for treating complex geometries.

We have applied the formalism to an exactly soluble model for 
nonresonant Fermi acceleration and found a transit-time correction 
factor that becomes exponentially small in the high-frequency limit.

Finally, we have illustrated these concepts using parameters from a 
fairly typical electron-cyclotron heating experiment.

\section*{Acknowledgments}

One of us (RLD) takes pleasure in acknowledging useful comments from 
Drs.  M.A. Lieberman, G.G. Borg, T.E. Sheridan, M.A. Tereschenko and 
S. Sridhar.  Numerical work and graphics was done using 
\emph{Mathematica 3.0} \cite{wolfram96}.


\begin{thebibliography}{10}

\bibitem{vedenov-velikhov-sagdeev62}
A.~A. Vedenov, E.~P. Velikhov, and R.~Z. Sagdeev, Nucl. Fusion Suppl. {\bf Pt.
  2},  465  (1962).

\bibitem{vedenov63}
A.~A. Vedenov, Plasma Phys. {\bf 5},  169  (1963).

\bibitem{drummond-pines62}
W.~E. Drummond and D. Pines, Nucl. Fusion Suppl. {\bf Pt. 3},  1049  (1962).

\bibitem{kaufman72}
A.~N. Kaufman, Phys. Fluids {\bf 16},  1063  (1972).

\bibitem{hazeltine-mahajan-hitchcock81}
R.~D. Hazeltine, S.~M. Mahajan, and D.~A. Hitchcock, Phys. Fluids {\bf 24},
  1164  (1981).

\bibitem{shurygin-dewar95}
R.~V. Shurygin and R.~L. Dewar, Plasma Phys. Control. Fusion {\bf 37},  1311-
  (1995).

\bibitem{stix92}
T.~H. Stix, {\em Waves in Plasmas} (American Institute of Physics, New York,
  1992).

\bibitem{chirikov79}
B.~V. Chirikov, Phys. Rep. {\bf 52},  263  (1979).

\bibitem{greene79}
J.~M. Greene, J. Math. Phys. {\bf 20},  1183  (1979).

\bibitem{rechester-rosenbluth-white79}
A.~B. Rechester, M.~N. Rosenbluth, and R.~B. White, Phys. Rev. Lett. {\bf 42},
  1247  (1979).

\bibitem{lichtenberg-lieberman92}
A.~J. Lichtenberg and M.~A. Lieberman, {\em Regular and Chaotic dynamics},
  No.~38 in {\em Applied Mathematical Sciences}, 2nd ed. (Springer, New York,
  1992), 1st ed. entitled Regular and Stochastic Motion (1983).

\bibitem{yoshida-etal98}
Z. Yoshida, H. Asakura, H. Kakuno, K.~T. J.~Morikawa, S. Takizawa, and T.
  Uchida, Phys. Rev. Lett. {\bf 12},  2458  (1998).

\bibitem{burke-pelletier92}
R. Burke and J. Pelletier,  in {\em Microwave Excited Plasmas}, {\em Plasma
  Technology, 4}, edited by M. Moisan and J. Pelletier (Elsevier, Amsterdam,
  1992), pp.\ 273--302.

\bibitem{lieberman-lichtenberg94}
M.~A. Lieberman and A.~J. Lichtenberg, {\em Principles of Plasma Discharges and
  Materials Processing} (Wiley, New York, 1994).

\bibitem{sinai70}
Y.~G. Sinai, Russian Math. Surveys {\bf 25},  137  (1970), russian orig.: Usp.
  Mat. Nauk XXV No. 2, 141--192 (1970).

\bibitem{ott93}
E. Ott, {\em Chaos in dynamical systems} (Cambridge Univ. Press, Cambridge,
  U.K., 1993).

\bibitem{dahlqvist-russberg90}
P. Dahlqvist and G. Russberg, Phys. Rev. Lett. {\bf 65},  2837  (1990).

\bibitem{ciubotariu-golovanivsky-bacal96}
C.~I. Ciubotariu, K.~S. Golovanivsky, and M. Bacal,  in {\em ICPP96: Proc. 1996
  International Conference on Plasma Physics}, edited by H. Sugai and T.
  Hayashi (The Japan Society of Plasma Science and Nuclear Fusion Research,
  Nagoya, Japan, 1997), Vol.~2, p.\ 1238.

\bibitem{ciubotariu97}
C.-I. Ciubotariu, D.Sc. thesis, Universit\'{e} Paris-Sud, 1997.

\bibitem{fermi49}
E. Fermi, Phys Rev. {\bf 75},  1169  (1949).

\bibitem{jarzynski94}
C. Jarzynski, Physica D {\bf 77},  276  (1994).

\bibitem{lieberman-godyak98}
M.~A. Lieberman and V.~A. Godyak, IEEE Trans. Plasma Sci. {\bf 26},  955
  (1998).

\bibitem{schmidt66}
G. Schmidt, {\em Physics of high temperature plasmas, an introduction}, 1st ed.
  (Academic Press, New York, 1966).

\bibitem{morse-feshbach53}
P.~M. Morse and H. Feshbach, {\em Methods of Theoretical Physics} (McGraw-Hill,
  New York, 1953).

\bibitem{greene-etal81}
J.~M. Greene, R.~S. MacKay, F. Vivaldi, and M.~J. Feigenbaum, Physica D {\bf
  3},  468  (1981).

\bibitem{dewar73}
R.~L. Dewar, Phys. Fluids {\bf 16},  1102  (1973).

\bibitem{wolfram96}
S. Wolfram, {\em The Mathematica Book}, 3rd ed. (Wolfram Media/Cambridge
  University Press, Champagne, Illinois, 1996).

\end{thebibliography}
\end{document}